\documentclass{article}
\usepackage{spconf,amsmath,graphicx}
\usepackage{multirow}

\title{Blind Omnidirectional Image Quality Assessment: Integrating Local Statistics and Global Semantics}
\name{Wei Zhou and Zhou Wang}
\address{Department of Electrical \& Computer Engineering, University of Waterloo, Canada\\
Email: \{wei.zhou, zhou.wang\}@uwaterloo.ca}
\begin{document}
%\ninept
%
\maketitle
\begin{abstract}
Omnidirectional image quality assessment (OIQA) aims to predict the perceptual quality of omnidirectional images that cover the whole 180$\times$360$^{\circ}$ viewing range of the visual environment. Here we propose a blind/no-reference OIQA method named S$^2$ that bridges the gap between low-level statistics and high-level semantics of omnidirectional images. Specifically, statistic and semantic features are extracted in separate paths from multiple local viewports and the hallucinated global omnidirectional image, respectively. A quality regression along with a weighting process is then followed that maps the extracted quality-aware features to a perceptual quality prediction. Experimental results demonstrate that the proposed S$^2$ method offers highly competitive performance against state-of-the-art methods.
\end{abstract}
\section{Introduction}
The rapid recent advancement in virtual reality (VR) technologies makes it possible to create immersive multimedia quality-of-experience (QoE) for end-users. As a representative form of VR, omnidirectional content has increasingly emerged in our daily life. To evaluate and optimize the perceptual QoE of omnidirectional content, objective omnidirectional image quality assessment (OIQA) models play a critical roles in the development of modern VR systems.

In the literature, objective OIQA models have emerged that follow both full-reference (FR) and no-reference (NR) frameworks. FR-OIQA models assume full access to information of the reference image and are usually direct extensions of traditional FR methods developed for regular rectangular 2D image quality assessment (IQA). For example, based upon the peak signal-to-noise ratio (PSNR), Yu et al. \cite{yu2015framework} propose the spherical PSNR (S-PSNR) algorithm, where PSNR is calculated for uniformly distributed points on a sphere instead of projected rectangular image. In \cite{sun2017weighted}, the weighted-to-spherically uniform PSNR (WS-PSNR) method is presented, where a weighting map is created by considering the stretched degree. Zakharchenko et al. \cite{zakharchenko2016quality} propose the Craster parabolic projection PSNR (CPP-PSNR) approach, which maps the reference and distorted omnidirectional images on the Craster parabolic projection followed by PSNR computation.

\begin{figure}[t]
	\centerline{\includegraphics[width=9.0cm]{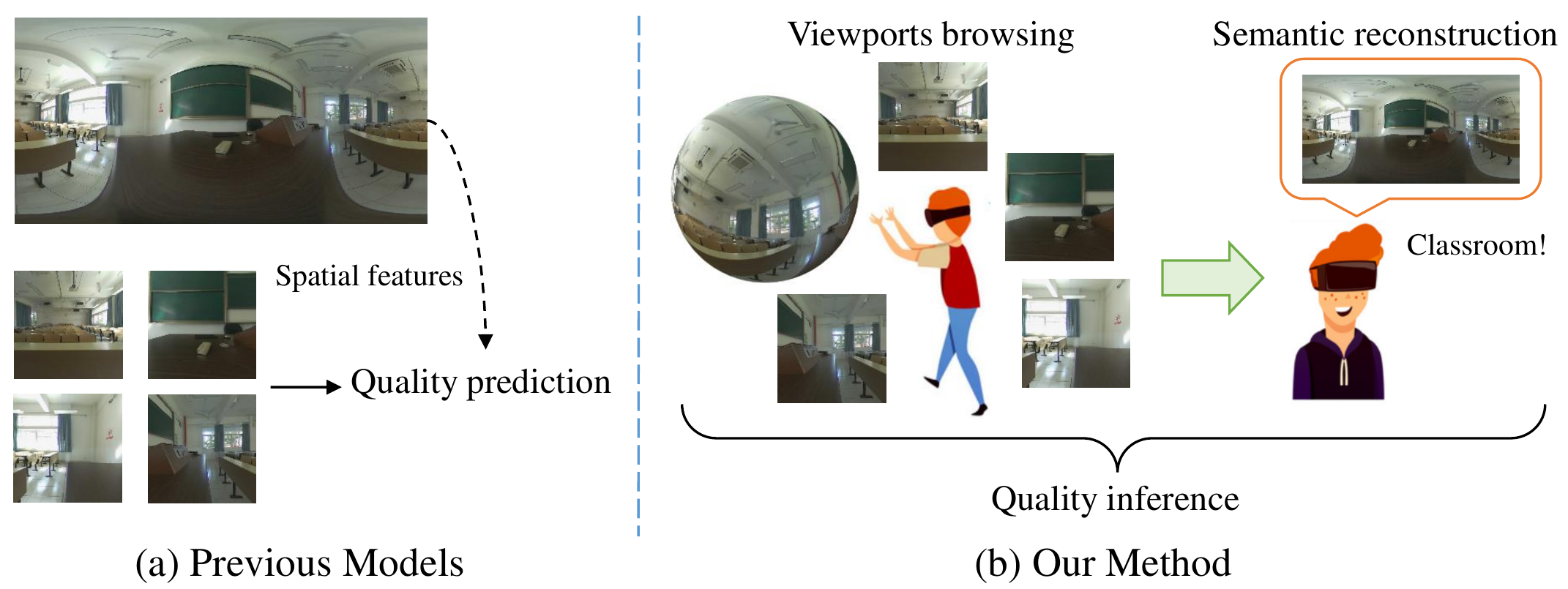}}
	\caption{Perceptual cues in omnidirectional image quality assessment. Existing models extract spatial information from various viewports and may obtain help from global projected maps, whereas the proposed method combines local image statistics and global semantic reconstruction.}
	\centering
	\label{fig1}
\end{figure}

\begin{figure*}[t]
	\centerline{\includegraphics[width=15.7cm]{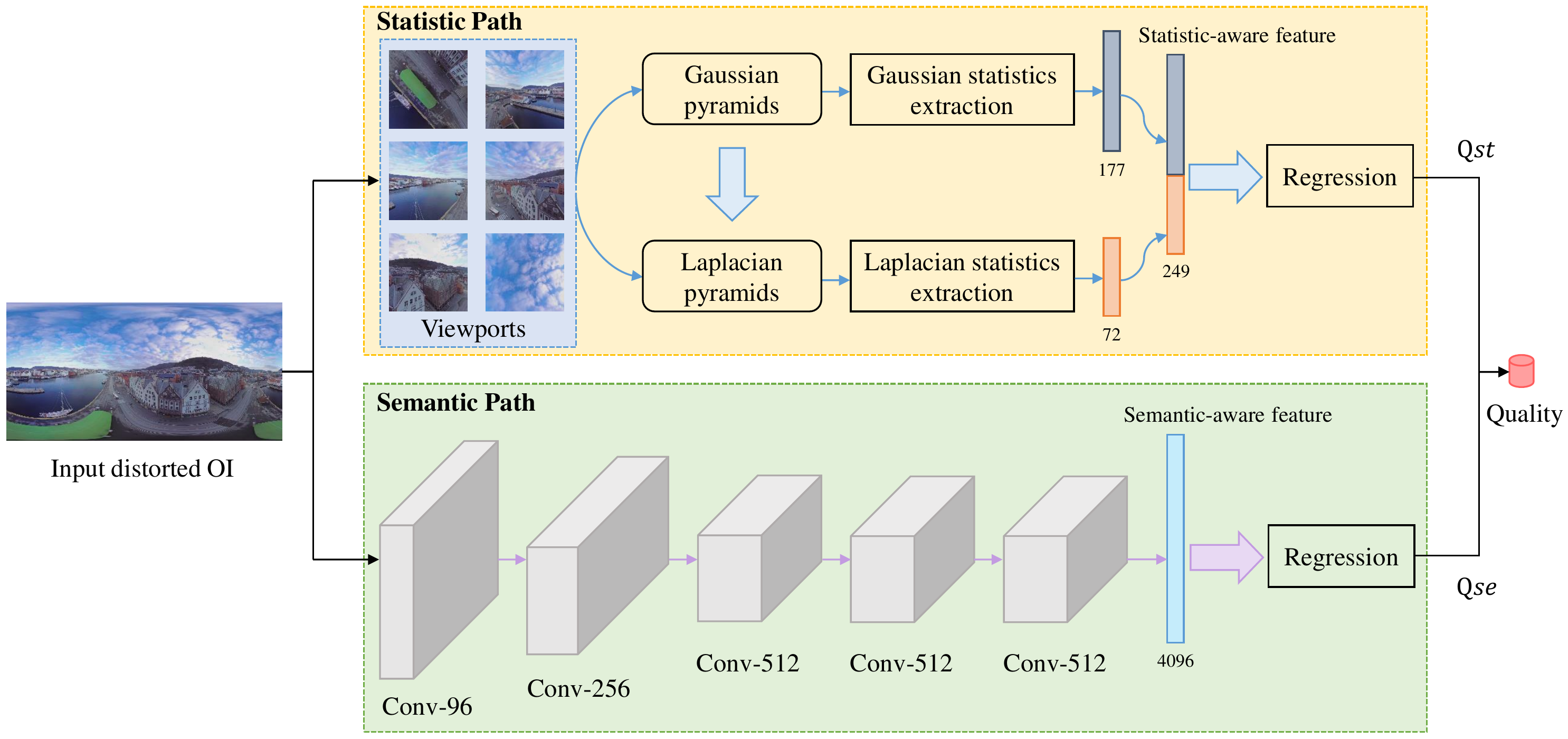}}
	\caption{Framework of the proposed S$^2$ method for blind OIQA.}
	\centering
	\label{fig2}
\end{figure*}

NR-OIQA methods do not require access to the reference image and are more desirable in many application scenarios. Existing NR-OIQA approaches can generally be classified into two categories, depending on whether the conventional hand-crafted or learned deep features are employed for quality prediction. Multi-frequency information and local-global naturalness are applied to develop the MFILGN model \cite{zhou2021no}. More recent models employ deep convolutional neural networks (CNNs) or graph convolution networks (GCNs). These models demonstrate promising performance, including the multi-channel CNN for blind 360-degree image quality assessment (MC360IQA) \cite{sun2019mc360iqa}, the viewport oriented graph convolution network (VGCN) \cite{xu2020blind}, and its variant named adaptive hypergraph convolutional network (AHGCN) \cite{fu2021adaptive}.

In a 360-degree viewing environment, e.g. using a head-mounted device, the observer is not able to visualize the whole omnidirectional content simultaneously, and thus an important step in the human subjective viewing experience is to establish or reconstruct a sense of the global semantics by browsing and integrating information from many viewports. During the course of image quality assessment, such global semantics are integrated with local observations on image fidelity, naturalness, and/or artifacts to produce an overall quality evaluation. Motivated by this observation, we propose a statistic and semantic oriented quality prediction framework named S$^2$ for blind OIQA as illustrated in Fig.~\ref{fig1}, by integrating features extracted from both low-level image statistics of multiple local viewports and high-level semantics of the hallucinated global omnidirectional image. A quality regression module is then leveraged to map the collection of the quality-sensitive features extracted from the two separate paths to an overall prediction of the subjective quality rating. Extensive experimental results demonstrate that the proposed method is superior to many state-of-the-art quality assessment models. In addition, we make some interesting observations on the relationship between semantic confidence and image distortions, as well as how the individual components affect the ultimate quality prediction performance in ablation studies.

\section{Proposed Method}
The overall framework of the proposed S$^2$ method is shown in Fig. \ref{fig2}, which consists of a statistic path, a semantic path, and a final quality regression step.
%since we do not rely on the reference information, the distorted omnidirectional image (OI) is used as the input.

%\subsection{Statistic Path}
Since a variety of viewports are browsed by the viewers, we first convert the distorted omnidirectional image (OI) to multiple viewports. Given each input distorted OI denoted by $D$, we exploit the non-uniform viewport sampling strategy \cite{xu2019quality,chen2020stereoscopic} and obtain $N$ viewports $V_n, n = 1, 2, ..., N$.

To capture the multi-scale characteristics of the human visual system \cite{wang2003multiscale}, we construct pyramid representations \cite{wang2005reduced,laparra2016perceptual} of multiple local viewports. %Based on empirical experiments, the performance cannot be further improved after the second level of Laplacian pyramids. Therefore, we choose the two-level Laplacian pyramids in our framework. To operate viewports using the two-level Laplacian pyramids, the three-time Gaussian pyramids should be first applied with the combination of low-pass filtering and 2 times down-sampling.
Specifically, multi-level Laplacian pyramids \cite{burt1987laplacian} are created by iterative Gaussian filtering, down-sampling, and subtracting, resulting in Gaussian and Laplacian pyramids in the same process. For a specific viewport $V_n$, layers of the Gaussian pyramid are calculated as follows:
%\begin{equation}
%\small
%G_{n}^{i}(x, y)=\left\{\begin{array}{l}
%V_{n}, i=1 \\
%\sum\limits_{u =  - 2}^2 {\sum\limits_{v =  - 2}^2 k(u, v) G_{n}^{i-1}(2 x+u, 2 y+v)}, i=2,3
%\end{array}\right.,
%\end{equation}
\begin{equation}
\small
G_{n}^{i}(x, y)=\left\{\begin{array}{l}
V_{n}, i=1 \\
\sum\limits_{u =  - 2}^2 {\sum\limits_{v =  - 2}^2 k(u, v) G_{n}^{i-1}(2 x+u, 2 y+v)}, i>1
\end{array}\right.,
\end{equation}
where $i$ is the layer index of the Gaussian pyramid, $x \in[0, X)$ and $y \in[0, Y)$ are the pixel position indices in which $X$ and $Y$ are the image dimensions, and $k(u, v)$ denotes the generating kernel that is typically defined by the coefficients of a low pass filter such as a 2D Gaussian filter.

We then interpolate each layer of the Gaussian pyramid by:
\begin{equation}
\small
\hat{G}_{n}^{i}(x, y)=4 \sum_{u=-2}^{2} \sum_{v=-2}^{2} k(u, v) G_{n}^{i}\left(\frac{u+x}{2}, \frac{v+y}{2}\right).
\end{equation}
The residual between the current layer of the Gaussian pyramid and the interpolation result from the next layer defines the current layer of the Laplacian pyramid:
\begin{equation}
\small
L_{n}^{i}=G_{n}^{i}-\hat{G}_{n}^{i+1}.
\end{equation}
Since the computation of the $i$-th layer in the Laplacian pyramid requires the $(i+1)$-th layer of the Gaussian pyramid, the number of layers in the Laplacian pyramid is one less than that in the Gaussian pyramid.% For example, if the Gaussian pyramid computation stops at $i=3$, then we will end up with a 3-layer Gaussian pyramid and a 2-layer Laplacian pyramid.
%\begin{equation}
%\small
%L_{n}^{j}=G_{n}^{j}-\hat{G}_{n}^{j+1},
%\end{equation}
%where $j = 1, 2$ indicates the $j$-th level of Laplacian pyramids. $G_{n}^{j}$ and $\hat{G}_{n}^{j+1}$ are the current $j$-th level generated Gaussian maps and the next $(j+1)$-th level interpolated Gaussian maps, respectively.

To extract features from the Gaussian pyramid, we compute the default uniform local binary pattern (LBP) descriptors, resulting in 59 statistics for each Gaussian layer. When a 3-layer Gaussian pyramid is employed, this leads to 177 Gaussian pyramid features denoted by $f_{GP}$. For a Laplacian pyramid, motivated by the success of natural scene statistics (NSS) in IQA research \cite{mittal2012making,fang2014no,chen2017blind}, we extract mean subtracted and contrast normalized coefficients, leading to 36 features for each layer. When a 2-layer Laplacian pyramid is employed, this results in 72 Laplacian pyramid features denoted by $f_{LP}$.
%With Gaussian and Laplacian pyramids, we extract the local statistics from generated Gaussian and Laplacian maps separately. It should be noted that we perform the average operation for the Gaussian and Laplacian statistics among different viewports.
%First, due to the rich texture content in Gaussian maps, we use the default uniform local binary pattern (LBP) descriptors from different Gaussian maps with 3 kinds of resolutions to serve as the Gaussian statistics denoted by $f_{GP}$. Each resolution of Gaussian maps leads to 59-dimensional statistics. Thus, the dimension of $f_{GP}$ is 177. Second, motivated by the great success of natural scene statistics (NSS) for evaluating perceptual image quality \cite{mittal2012making,fang2014no,chen2017blind}, we also extract mean subtracted and contrast normalized coefficients from different Laplacian maps with 2 kinds of resolutions, which are regarded as the Laplacian statistics, i.e. $f_{LP}$. Inspired by \cite{mittal2012no}, the original image scale and downsampled resolution scale by a factor of 2 are employed, leading to a 36-dimensional feature for each resolution. Hence, the dimension of $f_{LP}$ is 72. At last,
The full statistic feature set $f_{st}$, one for each viewport, is obtained by concatenating the statistical features extracted from the Gaussian and Laplacian pyramids as:
\begin{equation}
\small
f_{s t}=\left[f_{G P}, f_{L P}\right].
\end{equation}

%\subsection{Semantic Path}
%According to the browsed viewports, the global OI could be imagined for helping to predict the perceptual quality. Therefore, we extract deep semantics of distorted omnidirectional images, which are complementary to the local statistics from different viewports.

We employ the VGGNet trained on the large ImageNet dataset \cite{deng2009imagenet} as the semantic feature extraction backbone, mainly for its simplicity and ability to capture image distortion-related representations \cite{zhang2018unreasonable}. In \cite{chatfield2014return}, three different structures of VGGNet have been proposed to balance between complexity and accuracy, namely fast VGG (VGG-F), medium VGG (VGG-M), and slow VGG (VGG-S). Each of them contains 5 convolutional (Conv) layers and 3 fully connected (FC) layers. The first two FC layers have 4,096 neurons, while the last one has 1,000 nodes indicating the 1,000 classes for image recognition. In our current implementation, we select the deep features from the first FC layer of VGG-M as our semantic feature set $f_{se}$:
\begin{equation}
\small
f_{s e}=F C_{1} (D).
\end{equation}

%\subsection{Quality Regression}
To learn the mapping from features to quality labels, we feed the statistic features and semantic features separately to support vector regression (SVR) models \cite{chang2011libsvm}, and denote the regressed statistic and semantic quality scores as $Q_{st}$ and $Q_{se}$, respectively. The overall quality score is calculated by a weighted average:
\begin{equation}
\small
Q_{\text {overall }}=w Q_{s t}+(1-w) Q_{\text {se }},
\end{equation}
where $w$ is a weighting factor that determines the relative importance of the statistic and semantic feature predictors.% Empirically, we set $w$ to 0.9, which indicates that the statistic features are more reliable in quality prediction.

\section{Validation}

\subsection{Experimental Setup and Performance Comparison}
We evaluate the proposed approach on the CVIQD subjective database \cite{sun2018large}, which is so far a relatively large and widely adopted database containing both omnidirectional images and their corresponding quality labels given by human subjects. It consists of 16 original images and 528 distorted images produced by three classic image or video coding technologies, namely JPEG, AVC, and HEVC. The subjective quality ratings in the form of mean opinion score (MOS) are rescaled to the range of [0, 100], for which a higher MOS represents better perceptual image quality.

%The SROCC reflects the prediction monotonicity, while the PLCC indicates the prediction linearity. Moreover, the RMSE usually shows the prediction accuracy.

To compare the performance of various IQA models, we take Spearman Rank-Order Correlation Coefficient (SROCC), Pearson Linear Correlation Coefficient (PLCC) and Root Mean Squared Error (RMSE) as the evaluation criteria. Before calculating the PLCC and RMSE, a 5-parameter logistic nonlinear fitting approach \cite{video2003final} is implemented to map the predicted quality into the subjective quality space.% as follows:
%\begin{equation}
%\small
%q(x)=\lambda_{1}\left\{\frac{1}{2}-\frac{1}{1+\exp \left[\lambda_{2}\left(x-\lambda_{3}\right)\right]}\right\}+\lambda_{4} x+\lambda_{5},
%\end{equation}
%where $x$ and $q_x$ represent the raw predicted quality and the mapped quality score after the nonlinear regression process, respectively. $\lambda_{1}, \ldots, \lambda_{5}$ are the parameters to be fitted.

%Followed by \cite{zhou2021no},
The database is randomly divided into 80\% data for training and the remaining 20\% data for cross-validation. In order to relieve the uncertainty in training/testing splitting, we repeat this random-splitting and cross-validation process 100 times and report the median performance.

%\subsection{Performance Comparison with State-of-the-arts}

\begin{table}[t]
	\begin{center}
		\caption{Performance comparisons of objective models.}
		\label{table1}
		\scalebox{0.93}{
			\begin{tabular}{|c|c|c|c|c|}
                \hline
				Types & Methods & SROCC & PLCC & RMSE \\ \hline
				\multirow{5}{*}{FR-IQA} & PSNR & 0.6239 & 0.7008 & 9.9599 \\
                & SSIM \cite{wang2004image} & 0.8842 & 0.9002 & 6.0793 \\
                & MS-SSIM \cite{wang2003multiscale} & 0.8222 & 0.8521 & 7.3072 \\
                & FSIM \cite{zhang2011fsim} & 0.9152 & 0.9340	& 4.9864 \\
                & DeepQA \cite{kim2017deep} & 0.9292 & 0.9375 & 4.8574 \\ \hline
				\multirow{3}{*}{FR-OIQA} & S-PSNR \cite{yu2015framework} & 0.6449 & 0.7083 & 9.8564 \\
				& WS-PSNR \cite{sun2017weighted} & 0.6107 & 0.6729 & 10.3283 \\
				& CPP-PSNR \cite{zakharchenko2016quality} & 0.6265 & 0.6871 & 10.1448 \\ \hline
                \multirow{3}{*}{NR-IQA} & BRISQUE \cite{mittal2012no} & 0.8180 & 0.8376 & 7.6271 \\
                & BMPRI \cite{min2018blind} & 0.7470 & 0.7919	& 8.5258 \\
				& DB-CNN \cite{zhang2018blind} & 0.9308 & 0.9356 & 4.9311 \\ \hline
				\multirow{5}{*}{NR-OIQA} & MFILGN \cite{zhou2021no} & 0.9670 & 0.9751 & 3.1036 \\
				& MC360IQA \cite{sun2019mc360iqa} & 0.9428 & 0.9429 & 4.6506 \\
				& VGCN \cite{xu2020blind} & 0.9639 & 0.9651 & 3.6573 \\
                & AHGCN \cite{fu2021adaptive} & 0.9623 & 0.9643 & 3.6990 \\
				& Proposed S$^2$ & \textbf{0.9710} & \textbf{0.9781} & \textbf{2.8945} \\ \hline
		\end{tabular}}
	\end{center}
\end{table}

The performance of the proposed algorithm is compared against state-of-the-art quality assessment models, including five FR-IQA, three FR-OIQA, three NR-IQA, and four NR-OIQA methods. The results are shown in TABLE \ref{table1}, where we observe that for FR-IQA metrics, the PSNR-based models are inferior to more advanced approaches such as structural (SSIM, MS-SSIM, FSIM) and deep learning (DeepQA) models. Somewhat surprisingly, the FR-OIQA methods do not help further improve upon FR-IQA approaches. By contrast, the NR-OIQA models show significant superiority over NR-IQA methods. This is likely due to their specific design to capture the characteristics of omnidirectional images. Among all metrics tested, the proposed S$^2$ method demonstrates highly competitive performance. % This is mainly because conventional 2D metrics cannot well consider the specific characteristics of omnidirectional images.

\begin{figure}[t]
	\centerline{\includegraphics[width=9.0cm]{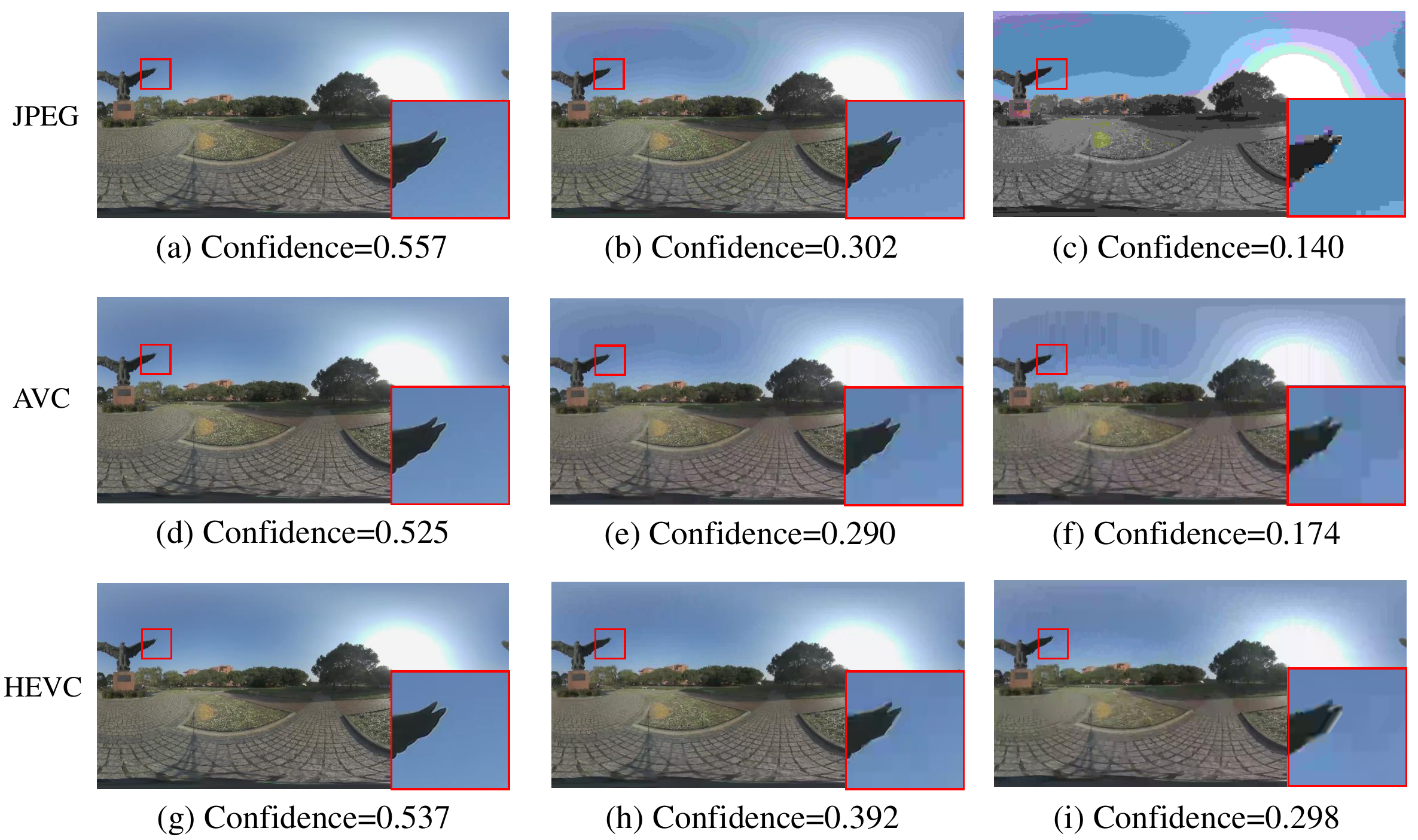}}
	\caption{The relationship between semantic confidence and image distortions. The first, second and third rows correspond to three distortion types (JPEG, AVC and HEVC compression, respectively). The first, second and third columns correspond to increasing distortion levels (low, medium and high, respectively) for each distortion type.}
	\centering
	\label{fig3}
\end{figure}

%\begin{table}[t]
%	\begin{center}
%		\caption{Performance results of ablation experiments. The best results are highlighted in bold.}
%		\label{table2}
%		\scalebox{1.2}{
%			\begin{tabular}{|c|c|c|c|}
%				\hline
%				Methods & SROCC & PLCC & RMSE \\ \hline
%				Statistic path & 0.9684 & 0.9769 & 3.0083 \\ \hline
%               Semantic path  & 0.9517 & 0.9576 & 4.0329 \\ \hline
%               Proposed S$^2$ & \textbf{0.9710} & \textbf{0.9781} & \textbf{2.8945} \\ \hline
%		\end{tabular}}
%	\end{center}
%\end{table}

\begin{figure}[t]
	\centerline{\includegraphics[width=7.0cm]{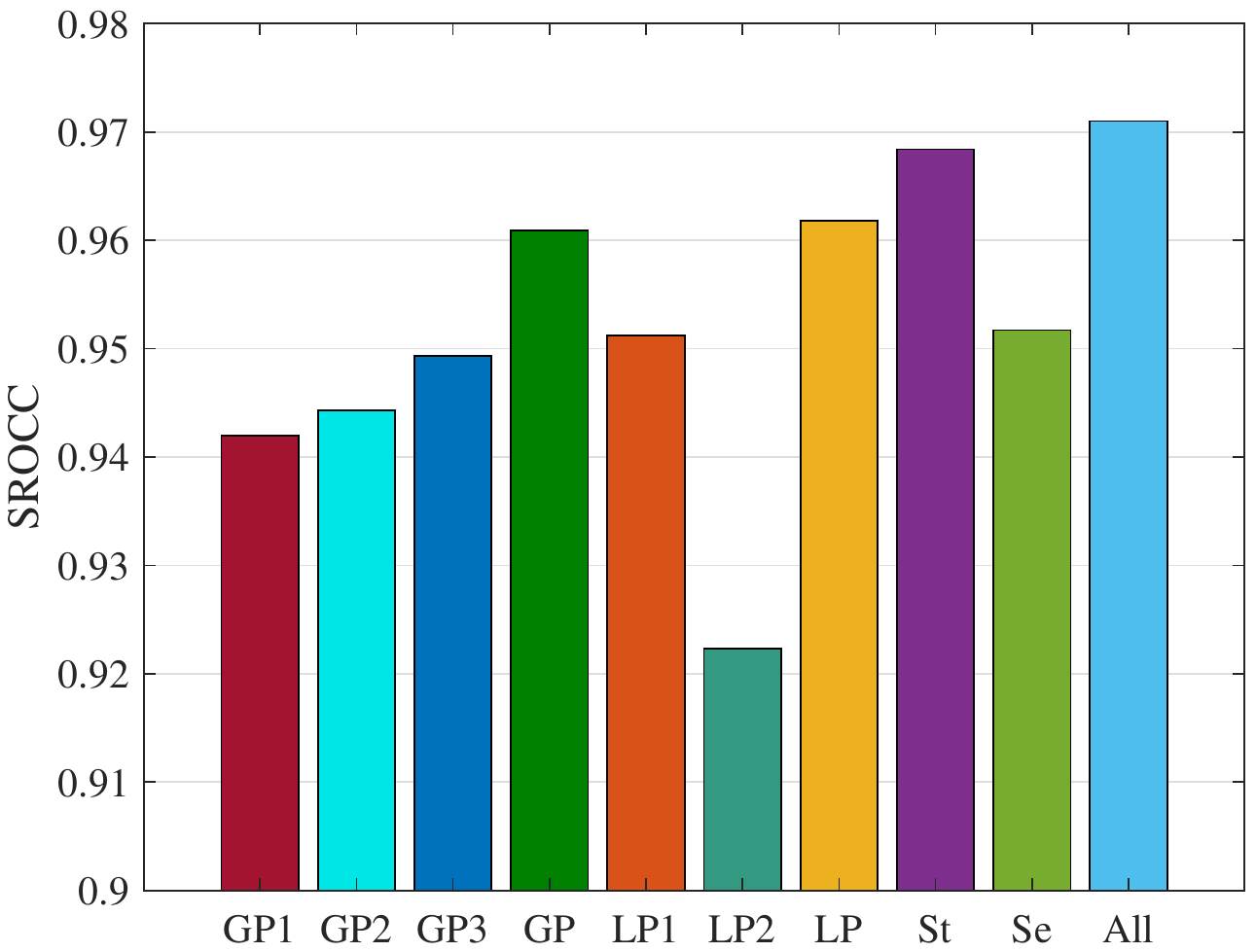}}
	\caption{Performance results of ablation experiments.}
	\centering
	\label{fig4}
\end{figure}

\subsection{Semantic Confidence Versus Image Distortion}
Since the proposed method contains a semantic path, it is interesting to observe the relationship between semantic confidence and image distortion. An example of distorted omnidirectional images with different JPEG, AVC and HEVC distortion levels is shown in Fig. \ref{fig3}, where from the first column to the third column, we observe that as the degree of distortion increases, the semantic confidence level decreases. This suggests that semantic information may be highly related to perceptual image quality. It is also interesting to see that the semantic confidence shows various sensitivities to different distortion types. In particular, the drop in semantic confidence levels is much less in more advanced image/video coding method HEVC than in earlier JPEG and AVC encoders. % If we compare the third column, i.e. those omnidirectional images with the most serious distortions, we observe that the HEVC compressed image has the highest semantic confidence. Therefore, we believe the proposed semantic clues provide a good criterion to different compression methods for omnidirectional images.

\subsection{Ablation and Parameter Sensitivity Tests}
We evaluate the contributions from the statistic and semantic paths by ablation experiments, and the results are shown in Fig. \ref{fig4}, where GP1, GP2 and GP3, respectively, represent the cases of using the first, second, and third layers of Gaussian pyramid statistics only. GP denotes the case of using three layers of Gaussian pyramid statistics. We find that the performance increases gradually. Similarly, LP1 and LP2, respectively, correspond to the cases of using the first and second layers of Laplacian pyramid statistics only, while LP denotes the case of using 2-layer Laplacian pyramid statistics. The results show that LP produces the best performance among the three. The cases of adopting the statistic path and the semantic path only are denoted by St and Se, respectively. It is observed that either path alone can achieve promising quality prediction performance, but adopting both paths (i.e. the All case) delivers the best performance. Relative speaking, the more dominant factor seems to be the statistic path. This may not be surprising as the statistic features come from different viewports directly visualized by human subject while the global semantics offer complementary information for additional cues in quality assessment.

\begin{table}[t]
	\begin{center}
		\caption{Performance comparisons for different viewport numbers in the statistic path.}
		\label{table2}
		\scalebox{0.93}{
			\begin{tabular}{|c|c|c|c|}
				\hline
				Numbers & SROCC & PLCC & RMSE \\ \hline
				$6$  & 0.9684 & 0.9769 & 3.0083 \\ \hline
                $20$ & 0.9686 & 0.9777 & 2.9626 \\ \hline
                $80$ & 0.9683 & 0.9771 & 2.9501 \\ \hline
		\end{tabular}}
	\end{center}
\end{table}

\begin{table}[t]
	\begin{center}
		\caption{Performance comparisons for different neural network architectures in the semantic path.}
		\label{table3}
		\scalebox{0.93}{
			\begin{tabular}{|c|c|c|c|}
				\hline
				Architectures & SROCC & PLCC & RMSE \\ \hline
				VGG-F & 0.9497 & 0.9537 & 4.2107 \\ \hline
                VGG-M & \textbf{0.9517} & \textbf{0.9576}& \textbf{4.0329} \\ \hline
                VGG-S & 0.9451 & 0.9486 & 4.4345 \\ \hline
		\end{tabular}}
	\end{center}
\end{table}

%\subsection{Parameter Sensitivity Tests}
Because different parameter settings may be employed in the implementations of the proposed framework, here we test the sensitivity of our model with regard to various viewport numbers and semantic architectures. The results are reported in TABLE \ref{table2} and TABLE \ref{table3}, respectively. We can see that the proposed model is insensitive to the viewport number. This allows us to reduce the number of viewports (for example, 6) to alleviate the computational complexity in real-world applications. The results also show that VGG-M outperforms the other neural network architectures in the semantic path. The possible reason may be that VGG-M achieves a preferable tradeoff between algorithm complexity and accuracy, making it a desired option for deep semantic backbone.

\section{Conclusion}
We propose a novel S$^2$ framework for blind omnidirectional image quality assessment that integrates both local low-level statistic and global high-level semantic features. Extensive experiments show that the proposed method achieves state-of-the-art performance. Observations on the relationship between semantic confidence and image distortion, and the ablation/sensitivity tests offer additonal useful insights. Under the same framework, more advanced models for statistic and semantic analysis may be employed in the future, aiming for more accurate QoE assessment models that may help drive the advancement of immersive multimedia systems.

\ninept
\bibliographystyle{IEEEbib}
\bibliography{strings,references}
\footnote{© 2023 IEEE. Personal use of this material is permitted. Permission from IEEE must be obtained for all other uses, in any current or future media, including reprinting/republishing this material for advertising or promotional purposes, creating new collective works, for resale or redistribution to servers or lists, or reuse of any copyrighted component of this work in other works.}
\end{document}